\documentclass[runningheads]{llncs}

\usepackage[T1]{fontenc}
\usepackage{graphicx}
\usepackage{hyperref}
\usepackage{amsmath}
\usepackage{xcolor}
\usepackage[backend=biber,style=numeric,sorting=ynt]{biblatex}
\begin{document}

\title{INSIGHT: A Survey of In-Network Systems for Intelligent, High-Efficiency AI and Topology Optimization}

\author{
Aleksandr Algazinov\textsuperscript{1}\and
Joydeep Chandra\textsuperscript{2} \and
Matt Laing\textsuperscript{3}
}

\institute{
\textsuperscript{1,2,3}Department of Computer Science and Technology, Tsinghua University, China \\
\email{algazinovalexandr@gmail.com}\textsuperscript{1}, \email{joydeepc2002@gmail.com}\textsuperscript{2}, \email{matthieu.laing@gmail.com}\textsuperscript{3}
}

\authorrunning{Aleksandr Algazinov\textsuperscript{1}\and
Joydeep Chandra\textsuperscript{2} \and
Matt Laing\textsuperscript{3}}
\titlerunning{INSIGHT: A Survey of In-Network Systems}

\maketitle

\begin{abstract}
In-network computation represents a transformative approach to addressing the escalating demands of Artificial Intelligence (AI) workloads on network infrastructure. By leveraging the processing capabilities of network devices such as switches, routers, and Network Interface Cards (NICs), this paradigm enables AI computations to be performed directly within the network fabric, significantly reducing latency, enhancing throughput, and optimizing resource utilization. This paper provides a comprehensive analysis of optimizing in-network computation for AI, exploring the evolution of programmable network architectures, such as Software-Defined Networking (SDN) and Programmable Data Planes (PDPs), and their convergence with AI. It examines methodologies for mapping AI models onto resource-constrained network devices, addressing challenges like limited memory and computational capabilities through efficient algorithm design and model compression techniques. The paper also highlights advancements in distributed learning, particularly in-network aggregation, and the potential of federated learning to enhance privacy and scalability. Frameworks like Planter and Quark are discussed for simplifying development, alongside key applications such as intelligent network monitoring, intrusion detection, traffic management, and Edge AI. Future research directions, including runtime programmability, standardized benchmarks, and new applications paradigms, are proposed to advance this rapidly evolving field. This survey underscores the potential of in-network AI to create intelligent, efficient, and responsive networks capable of meeting the demands of next-generation AI applications.
\end{abstract}


\section{Introduction}\label{sec1}
In-network computation, also known as in-network processing, represents how computational tasks are executed within a network infrastructure. Instead of relying on end devices or dedicated servers, in-network computation involves performing computations directly on the network devices, such as switches, routers, and Network Interface Cards (NICs). This approach enables applications to run within the network fabric, offering the potential for significantly higher throughput and lower latency compared to traditional models \cite{zheng2023innetwork}. AI workloads demand on network infrastructure, encompassing both model training and inference, often involve the processing of massive datasets. A recent survey \cite{zheng2023innetwork} indicated that over half of the data center experts indicate AI workloads to exert significant pressure on data center interconnect infrastructure in the future, surpassing even cloud computing. To meet these demands, a substantial portion of new data center facilities is expected to be dedicated to AI workloads, necessitating a re-evaluation of network architectures and capabilities. \\

Optimizing the development of in-network computation for AI is important for several reasons. Firstly, by performing AI-related computations closer to the data source and along the network path, in-network computation can substantially reduce latency, which is important for many AI applications, including real-time analytics, autonomous systems, and interactive services. Secondly, by processing data within the network, the volume of data that needs to be transmitted to centralized servers can be significantly reduced, removing bandwidth bottlenecks and improving overall network efficiency. In-network computation can enhance resource utilization by using the underutilized processing capabilities of network devices \cite{zheng2023innetwork}. \\

This paper contributes to the field of in-network computation for AI by providing a comprehensive analysis of the current state and future potential of integrating AI workloads with programmable network architectures. The survey explores the evolution of network paradigms, such as Software-Defined Networking (SDN) \cite{SDN} and Programmable Data Planes (PDPs) \cite{pdps}, detailing methodologies for mapping AI models onto resource-constrained network devices through efficient algorithm design and model compression techniques. The paper advances the understanding of distributed learning, particularly through in-network aggregation and federated learning, highlighting frameworks like Planter~\cite{zheng2024planter} and Quark \cite{quark} that simplify the development. By identifying key application areas such as intelligent network monitoring, intrusion detection, traffic management, and Edge AI and proposing future research directions like runtime programmability and standardized benchmarks, this survey serves as a resource for researchers and practitioners aiming to develop intelligent, efficient, and scalable network infrastructures for next-generation AI applications.

\section{Literature Review and Background}
 
\subsection{Evolution of Network Architectures and the Rise of Programmable Data Planes}
Traditional network devices, characterized by their fixed-functionality hardware, have long formed the backbone of internet infrastructure \cite{tanenbaum} . However, their rigid design often limited the ability to adapt to evolving network requirements and deploy innovative services \cite{michel2021programmable}. This closed-design paradigm, where functionalities were hard-coded by vendors, resulted in lengthy, costly, and inflexible processes for introducing new protocols or features \cite{michel2021programmable}. The increasing complexity of network traffic and the emergence of new applications, including AI, highlighted the need for greater flexibility and programmability within the network. The concept of Software-Defined Networking (SDN) \cite{SDN} is an approach to address these limitations. SDN fundamentally decouples the network's control plane, which dictates how traffic should be handled, from the data plane, which is responsible for the actual forwarding of packets. This separation allows for centralized control and management of network resources through a logically centralized controller, providing a platform for deploying new technologies and efficient algorithms \cite{quan2023ai}. \\

Building upon the foundation laid by SDN~\cite{SDN}, the development of Programmable Data Planes (PDPs) \cite{pdps} has great performance in network flexibility \cite{zheng2023innetwork}. PDP technology enables the systematic reconfiguration of the low-level processing steps applied to network packets~\cite{michel2021programmable}. This programmability allows network operators to customize packet forwarding logic and to execute advanced network functions directly within the data plane, leading to improved resource utilization \cite{zheng2023innetwork}. Modern network devices, such as switch Application-Specific Integrated Circuits (ASICs), NICs, and Field-Programmable Gate Arrays (FPGAs), leverage domain-specific languages like P4 to define and customize network protocols directly in the data plane. Using P4, operators can essentially instruct routers to conduct specified operations at terabit speeds for real-time decision-making and enhanced network visibility. This ability to flexibly reprogram the entire packet processing pipeline for offloading computations and applications to network devices is known as in-network computing.

\subsection{The Convergence of Artificial Intelligence and Networking}
The field of networking has adapted to Artificial Intelligence (AI) and its subset, machine learning (ML), to address a wide range of challenges. From traffic classification and anomaly detection to network configuration and resource management, ML algorithms are being employed to analyze vast amounts of network data, identify patterns, and make intelligent decisions to optimize network performance and efficiency. \\

The intersection of AI and networking has given rise to the concepts of "AI for Networking" and "Networking for AI"~\cite{zheng2023innetwork}. "AI for Networking"~\cite{ai4net} focuses on leveraging AI techniques to improve the design, operation, and management of network infrastructure. This includes using AI to enhance traffic shaping, load balancing, resource allocation, network optimization, and security~\cite{quan2023ai}. "Networking for AI," on the other hand, deals with designing and optimizing network architectures to efficiently support the unique demands of AI workloads, such as high bandwidth and low latency requirements~\cite{zheng2024planter}. \\

In-network computation represents a crucial convergence point between these two domains. It involves achieving "Networking for Machine Learning"~\cite{ai4net} by utilizing the network infrastructure to accelerate ML tasks and implementing "ML for networking" by embedding ML functionalities within the network to enhance its performance and capabilities~\cite{akem2024trick}.  

\subsection{Benefits and Challenges of In-Network Computation for AI}

Performing AI-related tasks within the network infrastructure offers several compelling advantages. One of the most significant benefits is the potential for reduced latency~\cite{zheng2023innetwork}. By processing data closer to its source and along the network path, in-network computation minimizes the delays associated with transmitting data to and from centralized processing units. This is important for latency-sensitive AI applications. Furthermore, in-network computation can lead to increased throughput by enabling parallel processing within the network and reducing the load on end hosts and servers \cite{ai4net}. The ability to perform computations within the network can also result in improved resource utilization by leveraging the processing capabilities of network devices that might otherwise remain idle \cite{zheng2023innetwork}. In the field of network security, in-network computation can serve as a "first line of defense" by enabling real-time anomaly detection and intrusion prevention directly within the network fabric. The distributed nature of in-network computation can facilitate the development of more distributed AI systems. \\

Despite these benefits, the development and deployment of in-network AI present several significant challenges. Programmable network devices, while offering increasing flexibility, have limited resources compared to traditional servers and GPUs. These limitations include constraints on memory (SRAM, TCAM), a restricted number of processing stages, and limited computational capabilities. Additionally, common ML data types like floating-point numbers are not natively supported in many data plane programming languages like P4, further complicating the implementation of complex AI models \cite{sapio2021scaling}. The basic arithmetic and logical operations natively supported by data plane programming languages might not be sufficient for certain AI algorithms, making complex calculations difficult \cite{zheng2023innetwork}. Balancing model accuracy with the limited resources and scalability of in-network AI solutions remains an ongoing challenge. Updating ML models deployed within the network without disrupting traffic flow is also a complex task. Furthermore, processing encrypted traffic for in-network AI is challenging as network devices typically lack decryption capabilities. The absence of unified benchmarks makes it difficult to compare different in-network AI solutions and assess their performance and resource consumption. Additionally, synchronization issues can arise in distributed in-network aggregation scenarios \cite{shi2024accelerating}. Overcoming these hurdles requires innovative approaches in algorithm design, model optimization, and resource management.

\section{Optimizing Development Methodologies for In-Network AI}

\subsection{Mapping Machine Learning Models to Programmable Data Planes}

A key step in enabling in-network AI is adapting trained ML models to operate within the architectural constraints of programmable network devices~\cite{gherari2023reviewinnetworkcomputingrole}. These devices typically impose limitations such as constrained memory, fixed pipeline stages, and minimal support for complex arithmetic operations. For decision tree-based models, two main approaches are used: the depth-based method, which maps each level of the tree to a stage in the processing pipeline, and the encode-based method, which transforms split criteria into binary encoded values to streamline and increase inference efficiency~\cite{Jamil_2022}. Different ML models require distinct deployment strategies. For instance, Binary Neural Networks (BNNs) \cite{yan2024brainonswitchadvancedintelligentnetwork}, which utilize binary weights and activations, can be effectively deployed on switches using lightweight bitwise operations such as XNOR and PopCount. Other approaches include direct mapping, which translates model logic into pipeline operations; encode-based mapping, which pre-processes features into discrete encodings; and lookup-based methods, which enable fast inference using precomputed tables. Lookup methods are effective for models like Naive Bayes and Support Vector Machines (SVMs) that rely heavily on repetitive calculations~\cite{cui2021netfcenablingaccuratefloatingpoint}. \\

Frameworks like Planter facilitate this transition by automatically compiling high-level ML models into switch executable instructions~\cite{liu2024operationalizingaifuturenetworks}. These tools support diverse algorithms, including classification, clustering, and dimensionality reduction, and assess feasibility using metrics such as pipeline utilization, inference latency, and model accuracy and efficiency.

\subsection{Addressing Resource Constraints through Efficient Algorithm Design}

Due to the resource constraints of network devices, such as limited memory, fixed instruction sets, and the absence of floating point support, ML models are expected to be lightweight and computation-friendly \cite{gherari2023reviewinnetworkcomputingrole}. Decision trees are well-suited to this resource-constrained environment, as they require only integer comparisons and minimal memory. In contrast, deeper architectures like conventional neural networks often go beyond hardware capabilities. To enable the proper functioning of more complex models, techniques such as fixed-point arithmetic and function linearization are employed~\cite{SwitchML}. Additionally, model partitioning across pipeline stages allows for limited parallelization. A promising strategy is hybrid inference \cite{Hybrid}, where lightweight models make quick decisions in-network, and more sophisticated inference occurs in servers. This architecture maintains high throughput while preserving inference and efficiency quality.

\subsection{The Role of Model Compression Techniques}

\begin{table}
\caption{Summary of Model Compression Techniques for In-Network AI}
\label{tab:compression}
\centering 
\tiny 
\setlength{\tabcolsep}{2pt} 
\begin{tabular}{|l|p{2.2cm}|p{2.0cm}|p{2.2cm}|p{2.4cm}|} 
\hline
\textbf{Compression Technique} & \textbf{Description} & \textbf{Benefits} & \textbf{Potential Drawbacks} & \textbf{Suitability for In-Network AI} \\
\hline
Pruning \cite{pruning} & Removing less important connections or weights from the network. & Reduced model size, potentially faster inference. & Accuracy loss if excessive pruning. & Highly suitable for reducing the number of parameters to fit within memory constraints. \\
\hline
Quantization \cite{quantization} & Reducing the precision of weights and activations (e.g., from float32 to int8). & Reduced model size, faster arithmetic operations, lower power consumption. & Accuracy loss if precision is too low. & Essential for overcoming the lack of floating-point support and reducing resource usage. \\
\hline
Low-Rank Factorization \cite{matrix_factor} & Approximating weight matrices with lower-rank matrices. & Reduced number of parameters, potentially faster matrix multiplication. & Can be computationally expensive to perform. & Useful for reducing the size of weight matrices in neural network layers. \\
\hline
Knowledge Distillation \cite{distillation} & Training a smaller model to mimic a larger, more accurate model. & Smaller student model with comparable performance to the teacher. & Requires training a larger teacher model first. & Can be effective in creating smaller, more efficient models suitable for network devices. \\
\hline
Neural Architecture Search \cite{NAS} & Automating the search for optimal model architectures. & Can discover more efficient architectures for specific tasks and platforms. & Computationally intensive search process. & Potential to find network-friendly architectures with reduced complexity. \\
\hline
\end{tabular}
\end{table}

Model compression is critical for deploying AI on resource-constrained network devices. As model sizes continue to increase, model compression ensures deployability without sacrificing performance~\cite{neill2020overviewneuralnetworkcompression}. Common methods include pruning \cite{pruning} (eliminating redundant weights), quantization \cite{quantization} (reducing precision), and low-rank factorization \cite{matrix_factor} (decomposing weight matrices). These techniques collectively reduce memory footprint and inference time. More advanced approaches include knowledge distillation \cite{distillation}, training a model to mimic a larger model’s behavior, and neural architecture search (NAS) algorithms~\cite{NAS}, which explore architectures optimized for specific hardware environments. These strategies yield compact yet effective models that operate within switch-level constraints. Emerging innovations like conditional computation and sparse attention \cite{sparse} dynamically activate only relevant portions of the model, further minimizing resource usage. When integrated with compression, these techniques significantly enhance the feasibility of deploying AI capabilities directly into the network data servers. \\

Table \ref{tab:compression} summarizes the descriptions, advantages, disadvantages, and suitability for in-network applications of these model compression techniques. The table emphasizes that while each method has unique benefits, like smaller models or more efficient computation, the drawbacks are observed as well, and the suitability of each method relies on performance needs and hardware limitations. 
 
\section{Advancements in Distributed Learning and In-Network Aggregation}

\subsection{Accelerating Distributed AI Training with In-Network Aggregation}

Training large-scale AI models often requires distributing the workload across multiple machines to handle extensive datasets and computational overload~\cite{SwitchML}. Data parallelism \cite{data_parallel} is a common strategy where each worker processes a subset of the training data and maintains a local model copy. Following each iteration, workers synchronize their model updates to ensure convergence, typically by grouping gradients across all nodes. This synchronization step is communication-intensive and becomes a bottleneck as model sizes and the number of workers scale \cite{SwitchML}. The repeated transmission of large parameter updates can saturate network capacity, leading to diminished scalability. In-network aggregation has emerged as a promising solution to this issue. By leveraging programmable network devices to perform gradient aggregation within the network itself, the communication burden on bottleneck issues can be significantly reduced, resulting in faster training, efficiency, and throughput.

\begin{figure}[h]
 \centering
 \includegraphics[width=0.85\linewidth]{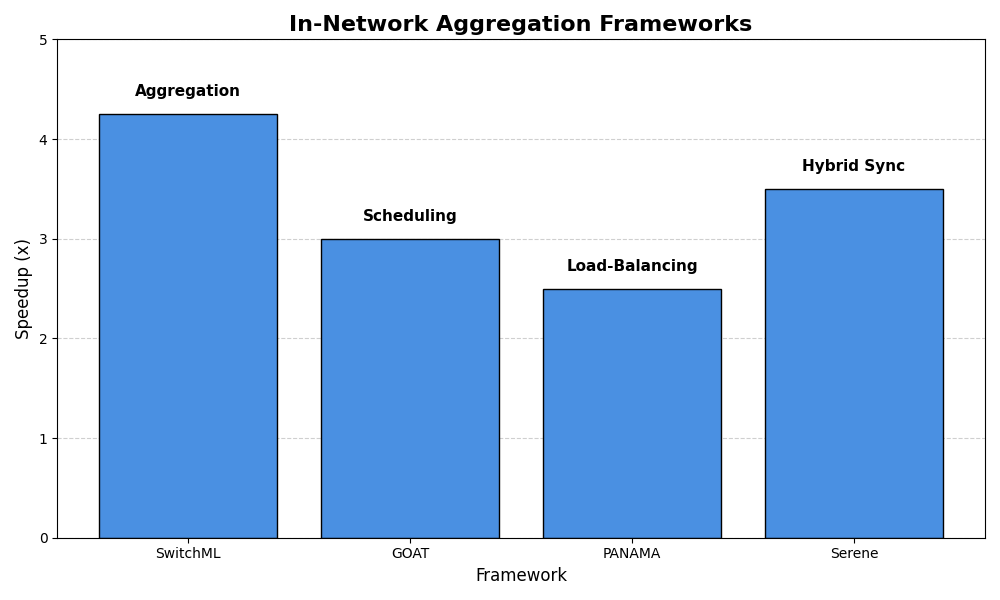}
 \caption{Comparison of Various In-Network Aggregations Frameworks}
 \label{fig:grids visualization}
\end{figure}

\subsection{Collaborative In-Network Aggregation Strategies and Frameworks}

Various frameworks were introduced to support collaborative in-network aggregation. A key challenge, however, lies in handling asynchronous gradient arrivals caused by network latency among workers. For example, GOAT \cite{goat} employs a scheduling mechanism that partitions models into sub-components and distributes aggregation across multiple switches, thereby accommodating asynchronous behavior \cite{FL}. PANAMA \cite{panama} integrates a load balancing algorithm and congestion control to fairly allocate bandwidth among distributed training jobs, while Serene \cite{serene} uses hybrid synchronization to mitigate straggler effects through coordination between programmable switches and ML clients. SwitchML is most used in academic research and experimental data center environments to accelerate distributed machine learning by leveraging in-network aggregation on programmable switches \cite{SwitchML}, which exploits programmable switches to aggregate model updates in the dataplane. As it can be seen in Fig. \ref{fig:grids visualization}, SwitchML performs the best in the speed up scenarios compared to the other frameworks like GOAT \cite{goat}, PANAMA \cite{panama}, and Serene \cite{serene}. SwitchML addresses hardware limitations, such as limited memory, absence of native floating point support, and constrained per-packet processing, by chunking model updates and applying quantization techniques to approximate floating point arithmetic. In addition, it introduces switch-side scoreboarding and end-host retransmissions to handle packet loss effectively. \\

Empirical evaluations have demonstrated significant improvements. For instance, SwitchML reduced training time by up to 3 times for deep neural networks and up to 5.5 times in more optimized scenarios \cite{SwitchML}. GOAT and Serene have also shown up to 40\% reductions in training completion time and improved synchronization efficiency, validating the effectiveness of these in-network aggregation frameworks.

\subsection{Federated Learning and its Potential in In-Network AI}

Federated Learning (FL) \cite{FL} is a decentralized approach to model training that allows numerous clients to collaborate without sharing their local data. In typical FL architectures, a central server coordinates training, and clients share only model updates. This preserves privacy and utilizes heterogeneous, decentralized data across devices such as smartphones or Internet of Things (IoT) sensors. Incorporating in-network computation into FL offers substantial performance enhancements. For instance, intermediate aggregation at edge nodes or programmable switches drastically reduces communication latency and server load. This hierarchical aggregation model accelerates convergence and scales better in distributed environments. \\

Centralized FL can face server-side bottlenecks, while model updates, though anonymized, can inadvertently leak private information. Therefore, mechanisms like differential privacy \cite{diff_privacy} and secure multiparty computation \cite{secure_mpc} remain necessary \cite{FL}. Other issues, such as data quality, device heterogeneity, computational constraints on edge devices, and privacy risks at intermediary nodes, must be addressed to fully realize the potential of the federated in-network AI.

\section{Frameworks and Tools for Simplifying In-Network AI Development}
Creating in-network computing AI applications can be challenging and imply expertise in both network hardware programming and ML. Numerous frameworks and tools have been developed to streamline the development process to reduce the barrier to entry, and enhance innovation in this area.

\begin{figure}[h]
  \centering 
  \includegraphics[width=0.85\linewidth]{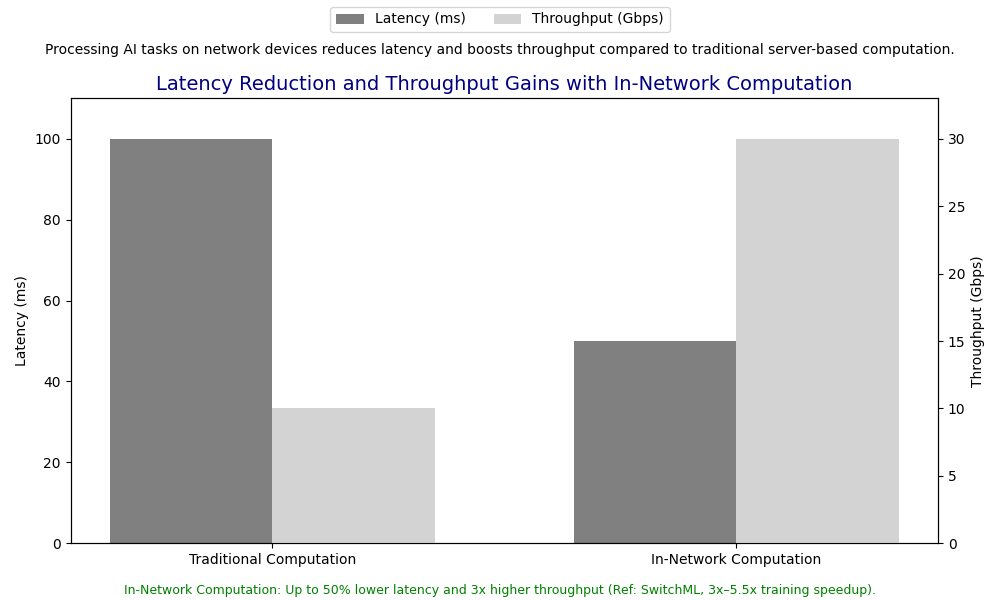}
  \caption{Latency Reduction and Throughput Gains with In-Network Computation}
  \label{fig:grids visualization2}
\end{figure}

\subsection{Overview of Existing Frameworks: Planter and Quark}

\begin{table}
\caption{Comparison of In-Network ML Frameworks}
\label{tab:frameworks}
\centering 
\tiny 
\setlength{\tabcolsep}{1pt} 
\begin{tabular}{|l|p{1.8cm}|p{2.2cm}|p{2.2cm}|p{1.8cm}|p{0.8cm}|p{0.8cm}|} 
\hline
\textbf{Framework Name} & \textbf{Primary Focus} & \textbf{Supported ML Models} & \textbf{Supported Hardware Platforms} & \textbf{Key Features} & \textbf{Ease of Use} \\
\hline
Planter \cite{zheng2024planter} & Rapid Prototyping of In-Network ML Inference & Tree-based (Decision Trees, Random Forests, XGBoost \cite{xgboost}, Isolation Forest \cite{isolation_forest}), Statistical (SVM, Naive Bayes, K-means, PCA), Neural Networks, Clustering & Intel Tofino/Tofino 2, AMe Alveo FPGA, NVIDIA Bluefield-2 DPU, P4Pi (BMv2, T4P4S)~\cite{P4}, DELL IoT Gateway & Modular Design, One-Click Deployment, Supports Multiple Architectures and Targets, Automated Development Lifecycle & High \\
\hline
Quark \cite{quark} & Fully Offloading CNN Inference onto PDPs & Convolutional Neural Networks (CNNs) & Intel Tofino ASIC (Hardware Switch), BMv2 (Software Switch) & Model Pruning, Quantization for Floating-Point Support, Modular CNN Design, Line-Line-Rate Inference with Low Latency & Medium \\
\hline
\end{tabular}
\end{table}

Planter \cite{zheng2024planter} is an open-source, modular framework designed for the rapid prototyping of in-network ML models across various platforms and pipeline architectures. The goal of Planter is to automate the overall process of creating, evaluating, and implementing in-network ML inference. Planter simplifies mapping trained models onto programmable data planes with little manual configuration by identifying general mapping methodologies for different ML algorithms. Several examples of supported ML models include tree-based models, statistical models, neural networks, clustering algorithms, and dimensionality reduction techniques. Planter is cross-platform, supporting a wide range of network hardware targets, including DELL IoT gateways, AMD Alveo FPGA, NVIDIA Bluefield-2 DPU, P4Pi (based on BMv2 and T4P4S)~\cite{P4}, and Intel Tofino and Tofino 2. New models, targets, architectures, and use cases can be easily integrated due to the specifics of the modular design. \\

Quark \cite{quark} is a framework specifically designed for fully offloading Convolutional Neural Network (CNN) inference onto Programmable Data Planes (PDPs) \cite{zheng2023innetwork}. While other frameworks and tools with the same name exist, they are used in different fields, such as a framework for data visualization or a framework for evaluating quantum algorithms. Our analysis refers only to Quark \cite{quark} in the context of in-network AI. By using model pruning to simplify CNN models and quantization techniques to support floating-point operations \cite{zheng2023innetwork}, it overcomes the resource constraints of PDPs \cite{pdps}. To maximize the use of resources on the PDP, Quark splits the CNN into smaller components. Quark achieves high accuracy in anomaly detection tasks while using a small fraction of the switch's resources and completing inference at line rate with low latency \cite{zheng2023innetwork}. This ability has been demonstrated by a tested prototype implemented on both P4 hardware switches (Intel Tofino ASIC) and software switches (BMv2). \\

Table \ref{tab:frameworks} provides a comparison of Planter and Quark, emphasizing their main objective, supported machine learning models, hardware platform compatibility, essential features, usability, and open-source status. Quark is optimized for low-latency CNN inference but has less clarity regarding accessibility and wider usability, whereas Planter supports a wider range of models and platforms with greater ease of use and open-source availability, making it more appropriate for rapid prototyping. 

\subsection{Modular Design and Rapid Prototyping of In-Network ML Models}
Rapid prototyping of in-network ML models is possible due to the modular design of frameworks such as Planter \cite{zheng2024planter}. Hence, various combinations of ML models, datasets, and target hardware platforms can be quickly tested by researchers and developers. To offload ML classification tasks into a programmable data plane, Planter, for instance, provides one-click capabilities. The required ML model and dataset are specified by users in a configuration file. Next, Planter is responsible for other parts of the pipeline, including training the model, converting it into a network device-compatible format, and creating the P4 code required for the target architecture. Faster iteration and experimentation are possible because of this automated process, which significantly lowers the time and expertise needed to deploy ML models in-network. Planter provides automated support for every stage of the development lifecycle, including training, deployment, and testing \cite{zheng2024planter}. Fig. \ref{fig:grids visualization2} shows that in-network computation can surpass the performance of the traditional server-based systems in both latency and throughput parameters.

\subsection{Cross-Platform Compatibility and Target Device Support}
Cross-platform compatibility is essential for in-network AI solutions to be useful and widely adopted. By supporting a wide variety of network hardware platforms, frameworks such as Planter satisfy this requirement \cite{zheng2024planter}. This enables users to test their in-network AI applications on various programmable devices, including software switches like BMv2, switch ASICs from vendors like Intel (Tofino series), FPGAs from AMD (Alveo series), and SmartNICs like NVIDIA Bluefield-2. These frameworks ensure a more open and accessible ecosystem for in-network AI research and development by supporting a variety of target architectures, ensuring that the developed solutions are not restricted to a particular hardware vendor or device type. This cross-platform capability is essential for the scalability and widespread deployment of in-network AI technologies in various networking environments. \\

\section{Applications of Optimize In-Network Computation for AI}
Optimization of in-network computation development for AI enables numerous applications in a variety of networking domains. Performance, efficiency, and security can be significantly increased for a variety of tasks by directly integrating intelligence into the network infrastructure. \\

\subsection{Intelligent Network Monitoring and Anomaly Detection} Intelligent network monitoring and anomaly detection are important applications of in-network AI \cite{ad_network}. By deploying ML models within switches and routers, traffic can be inspected and classified in real time. Hence, deviations from the typical behavior, such as congestion spikes, protocol misuse, or emerging threats, can be detected without the latency of sending data to external analyzers \cite{intelligent_network_op}. In particular, in-network ML has proven highly effective for identifying malicious bot traffic patterns with both low latency and over 90\% detection accuracy by processing features directly on the data plane \cite{botguard}. Frameworks like Quark \cite{quark}, which runs pruned and quantized convolutional neural networks entirely on programmable switches at line rate, and N-BaIoT \cite{n-baiot}, which employs deep autoencoders to flag anomalous IoT flows, have demonstrated sub-100 $\mu s$ inference times and high detection performance in real deployments. Beyond security, these capabilities can be leveraged for early-stage traffic classification to enable fine-grained Quality of Service (QoS) enforcement and dynamic load balancing within the network fabric \cite{QoS}.

\subsection{Enhanced Network Security through In-Network Intrusion Detection Systems} In-network computation is essential for improving network security by creating more efficient Intrusion Detection Systems (IDS) \cite{zheng2023innetwork}. Without using the control plane, line-rate inference can be used to detect and mitigate security threats by integrating AI-based security solutions straight into programmable network devices. This enables real-time network traffic analysis at wire speed, which helps identify anomalies and subtle attack patterns that conventional perimeter-based security solutions sometimes. Real-time intrusion detection benefits from the latency reduction. The reduction can be obtained by performing feature extraction at the data plane using languages such as P4~\cite{akem2024trick}. Hence, in-network IDS can effectively support real-time, wire-speed anomaly detection by performing feature extraction and AI-based inference directly at the data plane, thereby identifying subtle attack patterns with reduced latency without compromising network throughput.

\subsection{Optimized Traffic Management and Quality of Service} AI-optimized in-network computation is essential for dynamic and intelligent traffic management and Quality of Service (QoS) mechanisms \cite{zheng2023innetwork}. AI models can be embedded into network devices to make necessary decisions about resource allocation, traffic routing, and prioritization. These decisions are based on application requirements and real-time network conditions \cite{quan2023ai}. In addition, AI tools can be used to optimize forwarding rule storage, enhance content caching, and anticipate user needs for better load balancing. To increase throughput and guarantee the required bandwidth, in-network ML can help with dynamic network path selection and service type prioritization. The ability to perform early-stage traffic classification in the data plane allows for the embedding of Service Level Objectives (SLOs) into packet headers, enabling stateless QoS provisioning \cite{zheng2023innetwork}.

\subsection{Enabling Edge AI and Real-time Decision Making} In-network computation can be applied in Edge AI deployments, allowing AI inference to be performed closer to the data source, thereby significantly reducing latency. This is essential for AI systems that make real-time decisions, such as interactive AI services, industrial control systems, and autonomous vehicles \cite{autonomous_driving}. In-network AI reduces the latency involved in sending data to and from centralized cloud infrastructure. The reduction is accomplished by processing data at the network edge, in closer proximity to its generation point by sensors and devices. Furthermore, in-network computation can be advantageous for the effective aggregation of model updates in FL, which trains AI models on decentralized data at the edge.

\section{Future Outcomes and Research Directions}
The field of optimizing in-network computation for AI is rapidly evolving, and several promising future outcomes and research directions \cite{zheng2023innetwork} can be anticipated: \\

\begin{itemize}
    \item Runtime programmability. To adjust to shifting network conditions and evolving AI tasks, it is essential to enable dynamic updates of ML models and their parameters within the data plane without necessitating recompilation or disrupting traffic flow. \\

    \item Suitability for larger models. Optimizing in-network computation for large models is a viable and pertinent research area given the resource constraints of network devices. To satisfy the memory and processing power constraints, strategies like distributed in-network deployment, more sophisticated model compression, and model decomposition are crucial. \\

    \item Standardized benchmarks. To enable a fair comparison of various solutions and spur additional innovation in the field, it is important to develop new benchmarks and metrics for the performance assessment and resource consumption of various in-network ML algorithms across multiple platforms. \\

    \item New applications. There are numerous opportunities to investigate novel applications of in-network ML outside of the conventional networking domains, including computer vision, multimodal information processing, natural language processing, 6G networks, and smart cities.
\end{itemize}

\section{Conclusion}
Optimizing the development of in-network computation for AI represents an important approach to address the increasing demands of AI workloads on network infrastructure. By performing computational tasks directly within the network fabric, in-network AI offers the potential for significant reductions in latency, increased throughput, and improved resource utilization. This survey paper explored the evolution of existing network architectures that enabled this technology, the convergence of AI and networking, and the inherent benefits and challenges of in-network AI. Various methodologies for mapping AI models onto programmable data planes, techniques for addressing resource constraints through efficient algorithm design and model compression, and advancements in distributed learning and in-network aggregation were discussed. Frameworks like Planter and Quark are playing a crucial role in simplifying the development process. The applications of optimized in-network computation for AI are vast and include intelligent network monitoring, enhanced security through in-network intrusion detection, optimized traffic management, and the usage of Edge AI for real-time decision-making. While challenges remain, ongoing research and development in areas like runtime programmability, support for larger models, standardized benchmarks, and new use cases ensure that in-network AI plays a vital role in network infrastructure and the deployment of intelligent applications.


\printbibliography

\end{document}